\documentclass[a4paper,11pt]{article}
\usepackage{pos}

\title{On the Interplay of Nuclear and Higher-Twist Corrections in Nuclear Structure Functions}
\ShortTitle{On the interplay of nuclear and higher-twist corrections in nuclear structure functions}

\author[a]{S.I. Alekhin}
\author[b]{S.A. Kulagin}
\author*[c]{R. Petti}

\affiliation[a]{II. Institut f\"ur Theoretische Physik, Universit\"at Hamburg, D--22761 Hamburg, Germany}

\affiliation[b]{Institute for Nuclear Research of the Russian Academy of Sciences, 117312 Moscow, Russia}

\affiliation[c]{Department of Physics and Astronomy,
University of South Carolina, Columbia, SC 29208, USA}

\emailAdd{sergey.alekhin@desy.de}
\emailAdd{kulagin.physics@gmail.com} 
\emailAdd{roberto.petti@cern.ch}

\abstract{

We discuss results from our global QCD analyses including nuclear data
off deuterium from various measurements, as well as off  $\htri$ and $\hetri$ targets from the \mara{} experiment. We simultaneously determine the parton
distribution functions of the proton, the higher-twist terms, and the nucleon off-shell correction functions responsible
for the modifications of the partonic structure in bound protons and neutrons. 
In particular, we study the neutron-proton asymmetry of the off-shell correction and its interplay with the treatment of the higher-twist terms.
We observe that the data on the $\hetri/\htri, \hetri/\hdeu$, and $\htri/\hdeu$ cross section ratio are in good agreement with the predictions based on a single isoscalar off-shell function.
We provide the corresponding predictions on the ratio $F_2^n/F_2^p$, on the $d$ and $u$ quark distributions in the proton and in the $\htri$ and $\hetri$ nuclei, as well as for future measurements of the EMC effect with parity-violating Deep Inelastic Scattering (DIS). 

}

\FullConference{31st International Workshop on Deep Inelastic Scattering and Related Subjects (DIS2024)\\
 8–12 April 2024\\
Grenoble, France\\}

\newcommand{\mara}{{MARATHON}}

\newcommand{\ud}  {\mathrm{d}}

\newcommand{\gevsq}{\ \mathrm{GeV}^2}

\newcommand{\ceps}{\varepsilon}

\newcommand{\eq}[1]{Eq.~(\ref{#1})}
\newcommand{\Eqs}[2]{Eq.~(\ref{#1}) and (\ref{#2})}

\newcommand{\hdeu}{{}^2\text{H}}
\newcommand{\htri}{{}^3\text{H}}
\newcommand{\hetri}{{}^3\text{He}}

\usepackage{bm}

\begin{document}
\maketitle

\section{Introduction}
\renewcommand{\thefootnote}{\fnsymbol{footnote}}

The inclusion of data from deep-inelastic scattering off nuclear targets with different proton-neutron content in global QCD analyses allows to obtain insights on the physics mechanisms responsible of the modifications of bound nucleons in the nuclear environment, to accurately constrain the neutron PDFs, as well as to test the nucleon charge symmetry. 
In our recent analyses~\cite{Alekhin:2022tip,Alekhin:2022uwc}
we use deuterium DIS data from various experiments,
the data on the $\hetri/\htri$ cross section ratio from the \mara{} experiment~\cite{JeffersonLabHallATritium:2021usd},
along with a typical set of the proton DIS and collider data (for details see Refs.~\cite{Alekhin:2017fpf,Alekhin:2022tip}) to simultaneously constrain the proton PDFs, the higher-twist (HT) terms, and the functions describing the modification of the nucleon structure functions (SFs) in nuclei.
Nuclear corrections are treated following the microscopic model of Ref.~\cite{Kulagin:2004ie} addressing a number of effects relevant in different kinematic regions of Bjorken $x$.
The most important nuclear corrections in the large-$x$ region originate from the nuclear momentum distribution, the nuclear binding~\cite{Akulinichev:1985ij,Kulagin:1989mu} and the off-shell (OS) corrections to the bound nucleon SFs~\cite{Kulagin:1994fz,Kulagin:2004ie}.
The latter are directly related to the modification of the partonic structure of bound nucleons, as demonstrated in the analysis of data on the nuclear EMC effect~\cite{Kulagin:2004ie}.
The observations of Ref.~\cite{Kulagin:2004ie} have been confirmed in a global QCD analysis including deuterium DIS data~\cite{Alekhin:2017fpf,Alekhin:2022tip}.

The data from the \mara{} experiment on DIS cross sections off $\htri$ and $\hetri$ targets provides direct constraints on the nucleon isospin dependence of the OS functions~\cite{Alekhin:2022uwc} determining the in-medium modifications of the partonic structure of bound protons and neutrons. 
Since most of the existing fixed-target nuclear data have invariant momentum transfer squared $Q^2$ around few $\gevsq$ the HT contributions should be addressed.
To this end, we consider two different models for the HT terms and study their interplay with the resulting predictions on the ratio $d/u$ of the quark distributions, the structure function ratio $F_2^n/F_2^p$, and the proton-neutron asymmetry in the off-shell correction. Although we find that existing data are consistent 
with a single isoscalar off-shell function, we discuss how significant differences in the nuclear modifications of bound 
protons and neutrons can still be generated from conventional physics mechanisms.

\section{Theory and Analysis Framework}

The cross sections of the spin-independent charged-lepton inelastic scattering are fully described in terms of $F_T=2xF_1$ and $F_2$ SFs and $x$ is the Bjorken variable.
In the DIS region of high invariant momentum transfer squared $Q^2$,
SFs can be expressed as a power series in $Q^{-2}$ (twist expansion) within the operator product expansion (OPE): 
\begin{equation}\label{eq:sf}
F_i = F_i^{\text{TMC}} + H_i/ Q^2 + \cdots,  
\end{equation}
where $i=T,2$, $F_i^{\text{TMC}}$ are the corresponding leading twist (LT) SFs including the target mass correction (TMC)~\cite{Georgi:1976ve}, $H_i$ describes the twist-4 contribution.
The LT terms are given by a convolution of PDFs with the coefficient functions describing the quark-gluon interaction at the scale $Q$, which can be computed perturbatively as a series in the strong coupling constant (see, e.g.,~\cite{Accardi:2016ndt}).
We consider two models commonly used for the HT terms:
(i) additive HT model (aHT) motivated by the OPE, in which $H_i=H_i(x)$  and
(ii) multiplicative HT model (mHT)~\cite{Virchaux:1991jc},
in which $H_i$ is assumed to be proportional to the corresponding LT SF, $H_i=F_i^\text{LT}(x,Q^2) h_i(x)$.

Following Ref.~\cite{Kulagin:2004ie}, we address nuclear corrections in the DIS process at large $x$ by treating it as an incoherent scattering off bound nucleons in the target rest frame.
The deuteron SFs can be calculated as the sum of bound proton and neutron SFs convoluted with the  nucleon momentum distribution given by the deuteron wave function squared,
$\left|\Psi_d(\bm k)\right|^2$:
\begin{align}
	\label{eq:IA2}
	F_i^d = \int\!\ud^3\bm k K_{ij} \left|\Psi_d(\bm k)\right|^2 \left(F_j^p + F_j^n\right),
\end{align}
where the integration is performed over the bound nucleon momentum $\bm k$, $i,j=T,2$, we assume a summation over the repeated index $j$, and $K_{ij}$ are the kinematic factors~\cite{Kulagin:2004ie,Alekhin:2022tip}.

For nuclei with the number of nucleons $A\ge 3$
the convolution by \eq{eq:IA2} requires the integration over the energy spectrum of the residual nuclear system, along with the nucleon momentum, which are described by the nuclear spectral functions $\mathcal P_{p/A}$ and $\mathcal P_{n/A}$~\cite{Akulinichev:1985ij,Kulagin:1989mu,Kulagin:1994fz,Kulagin:2004ie,Kulagin:2010gd}:
\begin{align}
\label{eq:IA}
F_i^A = \int\!\ud^4 k K_{ij} \left(\mathcal{P}_{p/A} F_j^p + \mathcal{P}_{n/A} F_j^n\right),
\end{align}
where the integration is performed over the bound nucleon four-momentum $k$.
The proton (neutron) nuclear spectral function $\mathcal P_{p(n)/A}(\ceps,\bm k)$ describes the corresponding energy ($\ceps=k_0-M$) and momentum ($\bm k$) distribution in the target nucleus at rest and 
involves contributions from all possible $A-1$ intermediate states.

The proton (neutron) off-shell SFs in \eq{eq:IA2} and \eq{eq:IA} depend on the scaling variable $x'=Q^2/2k\cdot q$,
the DIS scale $Q^2$, and the nucleon invariant mass squared $k^2=k_0^2-\bm k^2\not=M^2$, where $M$ is the nucleon mass.
This latter dependence originates from both the power TMC terms of the order $k^2/Q^2$ and the OS dependence of the LT SFs.
Following Refs.~\cite{Kulagin:1994fz,Kulagin:2004ie},
we treat the OS correction in the vicinity of the mass shell $k^2=M^2$ by expanding SFs in a power series in $v=(k^2-M^2)/M^2$.
To the leading order in $v$ we have
\begin{align}
\label{SF:OS}
F_i^\text{LT}(x,Q^2,k^2) &= F_i^\text{LT}(x,Q^2,M^2)\left( 1+\delta f_i\,v \right),
\\
\label{eq:deltaf}
\delta f_i &= \partial \ln F_i^\text{LT}(x,Q^2,k^2)/\partial \ln k^2,
\end{align}
where the derivative is taken on the mass shell $k^2=M^2$.
We assume equal functions $\delta f_T=\delta f_2=\delta f$ for $F_T$ and $F_2$,
motivated by the observation that $F_T\approx F_2$ in the region for which the OS effect is numerically important~\cite{Kulagin:2004ie,Kulagin:2010gd,Alekhin:2017fpf,Alekhin:2022tip}.
The universal function $\delta f$ is independent from the nucleus considered and drives the nuclear modification of the bound nucleons in the nuclear environment. 

The details of \Eqs{eq:IA2}{eq:IA} can be found in Refs.~\cite{Kulagin:2004ie,Kulagin:2010gd,Alekhin:2022tip,Alekhin:2022uwc}.
Other nuclear effects like the meson-exchange currents and the nuclear shadowing 
result in corrections comparable to the 
experimental uncertainties at large $x$~\cite{Alekhin:2017fpf} and are therefore 
neglected in the present analysis. 

We simultaneously constrain the proton PDFs, the HT corrections, and the proton and the neutron OS functions, $\delta f^p$ and $\delta f^n$, describing the modifications the proton and neutron PDFs in the nuclear environment, in a global QCD analysis. 
We apply \Eqs{eq:IA2}{eq:IA} to address the nuclear corrections from the momentum distribution, the nuclear binding, and the OS effect, which are the main nuclear corrections at large $x$.
We use a deuteron wave function based on the Argonne nucleon-nucleon potential~\cite{Wiringa:1994wb,Veerasamy:2011ak} (AV18) as well as
the $^3$He and $^3$H spectral functions of Ref.~\cite{Pace:2001cm}
computed with the AV18 potential and accounting for the Urbana three-nucleon interaction, as well as the Coulomb effect in $\hetri$.
The datasets used are described in Refs.~\cite{Alekhin:2017fpf,Alekhin:2022tip} 
and include charged-lepton DIS data off proton, deuterium, $\htri$, and $\hetri$ targets, as well as data from the $W^\pm/Z$ boson production at hadron colliders.
In particular, data on the ratio of the DIS cross sections of the three-body nuclei, $\sigma^{\hetri}/\sigma^{\htri}$, $\sigma^{\hetri}/\sigma^{\hdeu}$, and $\sigma^{\htri}/\sigma^{\hdeu}$ from the \mara{} experiment~\cite{JeffersonLabHallATritium:2021usd,Abrams:2024wgt} allow to study the neutron-proton asymmetry  $\delta f^a=\delta f^n-\delta f^p$~\cite{Alekhin:2022uwc}. 
More detail on the analysis setup can be found in Refs.~\cite{Alekhin:2022tip,Alekhin:2022uwc}.

\section{Results and Discussion}
\begin{figure}[htb]
\centering
\includegraphics[width=1.00\textwidth]{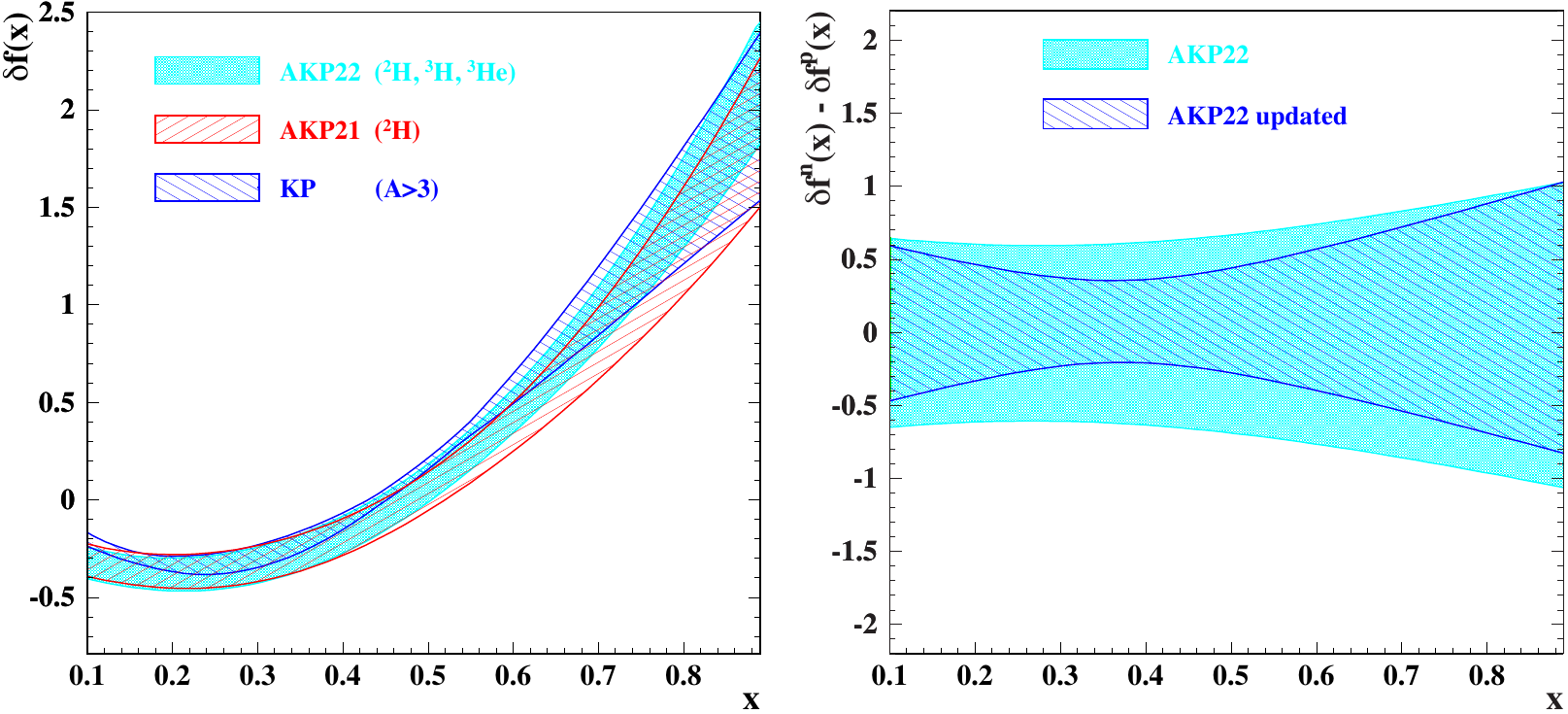}
\caption{%
Left:
The $1\sigma$  uncertainty band on the OS function obtained using $\hdeu,\htri,\hetri$ data  and assuming $\delta f^p=\delta f^n$ and the aHT model for the HT terms (AKP22 shaded area)~\cite{Alekhin:2022uwc}.
The results obtained from $A\ge 4$ (KP)~\cite{Kulagin:2004ie} and $A=2$ (AKP21) analyses~\cite{Alekhin:2022tip} are also shown for comparison.
Right:
The $1\sigma$ uncertainty band on the neutron-proton asymmetry $\delta f^n(x)-\delta f^p(x)$ obtained with the aHT model using 
two different data sets for the $A=3$ nuclei: 
(i) $\sigma^{\hetri}/\sigma^{\htri}$ data from Ref~\cite{JeffersonLabHallATritium:2021usd}; 
(ii) $\sigma^{\hetri}/\sigma^{\hdeu}$ data from Refs.~\cite{Abrams:2024wgt,Seely:2009gt,HERMES:1999bwb}, and $\sigma^{\htri}/\sigma^{\hdeu}$ data from Ref.~\cite{Abrams:2024wgt}. 
\label{fig:deltaf}} 
\end{figure}

In our QCD analysis we first assume equal off-shell functions for protons and neutrons, $\delta f^p=\delta f^n=\delta f$, and the aHT model for the HT terms. With such settings we obtain~\cite{Alekhin:2022uwc} a good agreement with the \mara{} data on the ratio $\sigma^{\hetri}/\sigma^{\htri}$~\cite{JeffersonLabHallATritium:2021usd}
with $\chi^2$ per number of data points (NDP) of $20/22$, and $\chi^2/\text{NDP}=4861/4065$ considering all data~\cite{Alekhin:2022tip,Alekhin:2022uwc}.

The function $\delta f(x)$ obtained from the analysis of $\hdeu,\htri,\hetri$ data of Ref.~\cite{Alekhin:2022uwc} is shown in Fig.~\ref{fig:deltaf} (left panel).
The results are in good agreement with the original determination of $\delta f(x)$~\cite{Kulagin:2004ie} from the ratios $\sigma^A/\sigma^d$ of the DIS cross sections
off nuclear targets with a mass number $A\geq 4$ and with those of Ref.~\cite{Alekhin:2022tip} based on nuclear data 
from deuterium ($A=2$) only. The consistency of results obtained from different nuclei is in line with the universality of the OS function. 

The validity of an assumption $\delta f^p=\delta f^n$ was verified in the analysis of the EMC effect in Ref.~\cite{Kulagin:2004ie} and the same assumption was also used in Refs.~\cite{Alekhin:2017fpf,Alekhin:2022tip,Kulagin:2010gd, Kulagin:2014vsa, Kulagin:2007ju, Ru:2016wfx}.
The \mara{} data on the ratio $\sigma^{\hetri}/\sigma^{\htri}$ were used to constrain the asymmetry $\delta f^a=\delta f^n-\delta f^p$~\cite{Alekhin:2022uwc}.
With the aHT model we obtain a function 
$\delta f^p$ similar to that of the isospin-symmetric case shown in Fig.~\ref{fig:deltaf} (left panel), as well as an asymmetry $\delta f^a$ consistent with zero within uncertainties, as shown in Fig.~\ref{fig:deltaf} (right panel)~\cite{Alekhin:2022uwc}.
To improve the sensitivity of our analysis to the asymmetry $\delta f^a$ we replace the $\sigma^{\hetri}/\sigma^{\htri}$ ratio from Ref.~\cite{JeffersonLabHallATritium:2021usd} with the recent measurements of 
$\sigma^{\hetri}/\sigma^{\hdeu}$ and $\sigma^{\htri}/\sigma^{\hdeu}$ from \mara{}~\cite{Abrams:2024wgt}, and also include the 
$\sigma^{\hetri}/\sigma^{\hdeu}$ measurements from the JLab Hall C~\cite{Seely:2009gt} and HERMES~\cite{HERMES:1999bwb} experiments.
The results are shown in Fig.~\ref{fig:deltaf} (right panel). While with the new $A=3$ data~\cite{Abrams:2024wgt} the asymmetry $\delta f^a$ is still consistent with zero, we obtain a significant reduction of the uncertainties compared with our former analysis of Ref.~\cite{Alekhin:2022uwc}. 

\begin{figure}[htb]
\centering
\includegraphics[width=1.00\textwidth]{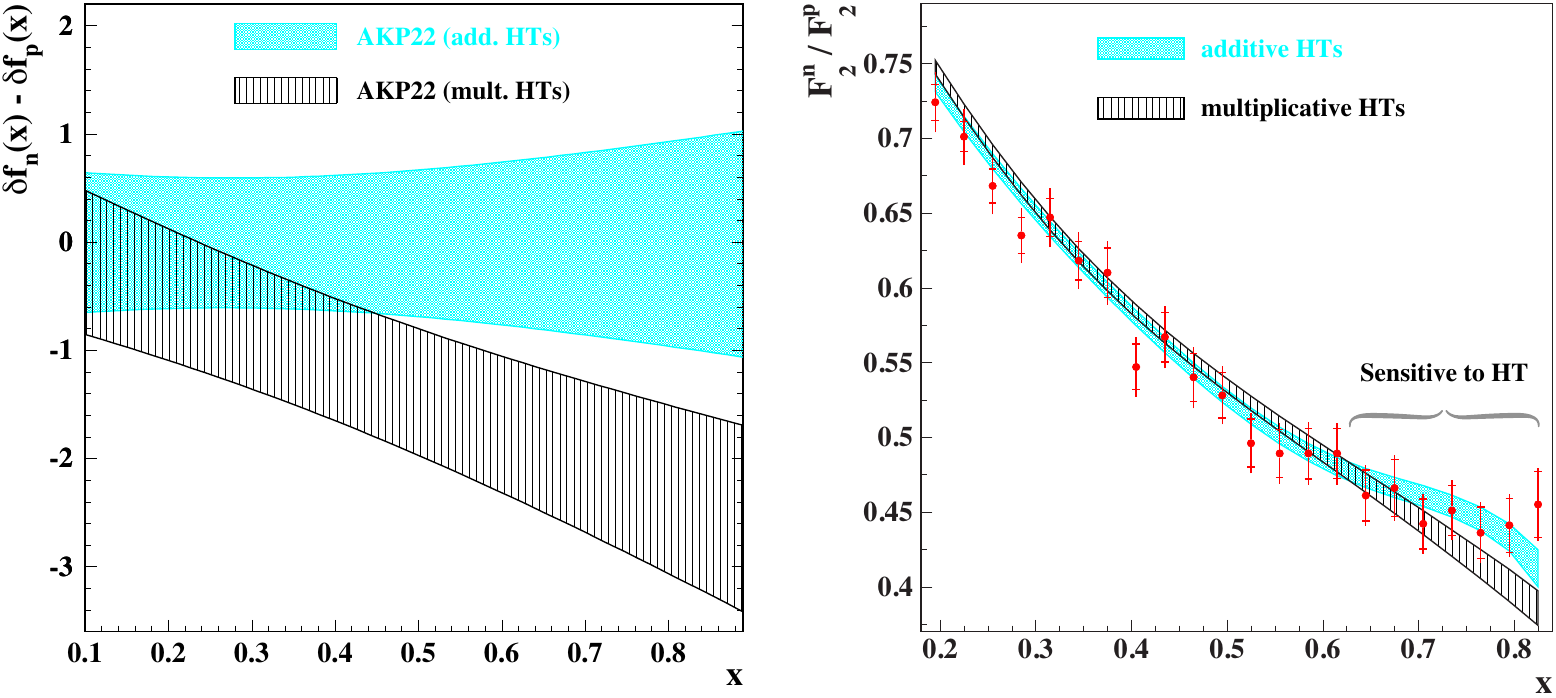}
\caption{%
Left:
The $1\sigma$ uncertainty band on the neutron-proton asymmetry $\delta f^n(x)-\delta f^p(x)$ for the aHT (shaded area) and mHT (hashed area) models~\cite{Alekhin:2022uwc}.
Right: Comparison of the \mara{} data on $F_2^n/F_2^p$~\cite{JeffersonLabHallATritium:2021usd} with the predictions ($1\sigma$ uncertainty band) obtained from Ref.~\cite{Alekhin:2022uwc} with the aHT model (shaded area) and the mHT model (hashed area).
\label{fig:HTmodel}} 
\end{figure}
We emphasize the importance of properly taking into account HT terms in the QCD analysis of DIS data with $Q^2 \lesssim 10$ GeV$^2$. 
While the aHT model provides a good description of data  with isoscalar HT terms and OS function, with the mHT model we 
observe various biases and inconsistencies as a result of the 
interplay between HT and LT terms inherent to the mHT model. 
In the mHT model we find a nonzero neutron-proton asymmetry in the OS function (left panel of Fig.\ref{fig:HTmodel}), directly correlated with an enhancement of the ratio $d/u$ at large $x$ compared to that in the aHT model.
These results are driven by the \mara{} $\hetri/\htri$ data,
whose description in the region $x >0.6$ is affected by the isospin effects in $\delta f$ as well as by the HT contributions.
Figure~\ref{fig:HTmodel} (right panel) shows a comparison of the \mara{} $F_2^n/F_2^p$ measurement with our predictions for both the aHT and the mHT models. Overall, we obtain an excellent description of the \mara{} data using a QCD analysis with the aHT model, with a $\chi^2/\text{NDP}=20/22$. 
The data clearly prefer the aHT model over the mHT model, as indicated by the higher $\chi^2/\text{NDP}=34/22$ with the latter. 

\begin{figure}[htb]
\centering
\includegraphics[width=1.00\textwidth]{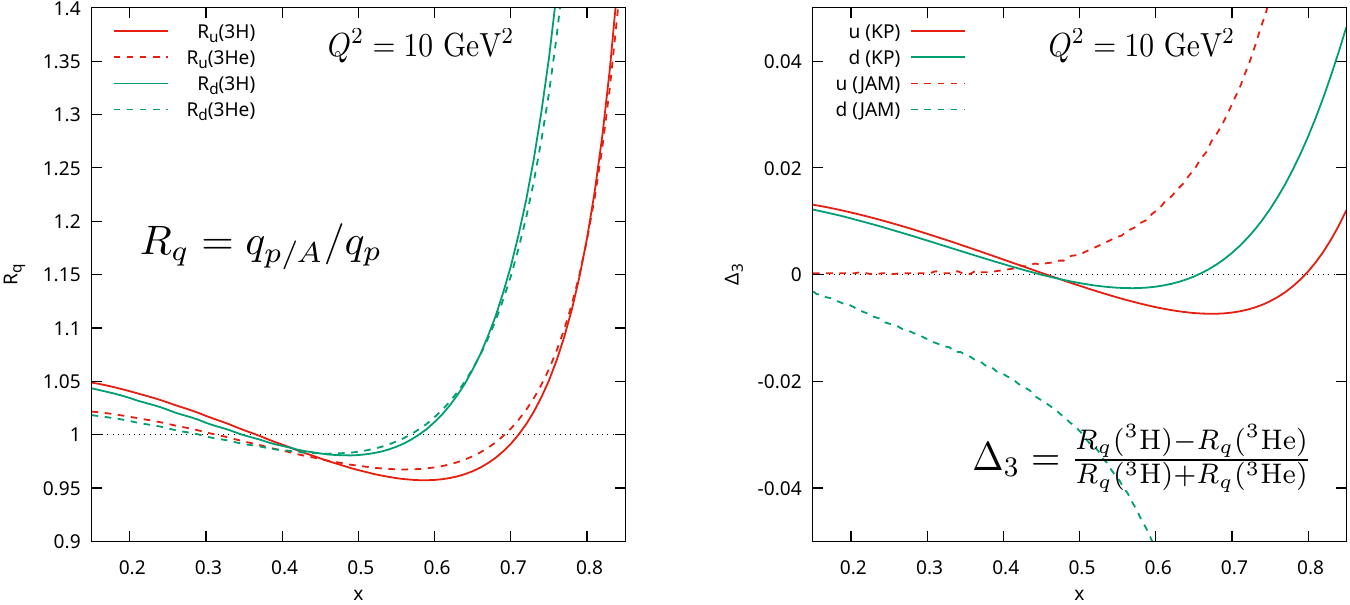}
\caption{%
Left:
The ratio $R_u$ and $R_d$ computed using the results of the analysis of Ref.~\cite{Alekhin:2022uwc} at $Q^2=10\gevsq$ for the $\htri$ and $\hetri$ nuclei.
Right:
The corresponding $\htri-\hetri$ asymmetry $\Delta_3$ for $u$ and $d$ valence quark distributions in bound proton.
\label{fig:RduA3}} 
\end{figure}
In order to further address the isospin effects in the nuclear corrections on various quark PDFs, we calculated the ratio $R_q=q_{p/A}/q_p$ 
of the proton contribution within the nucleus $A$, $q_{p/A}$, and the corresponding free proton PDF for both $u$ and $d$ quarks in $\hetri$ and $\htri$ using the proton PDFs and the $\delta f(x)$ function of Ref.~\cite{Alekhin:2022uwc}, shown in Fig.~\ref{fig:RduA3}.
The ratio $R_q$ describes the modifications of the parton distributions $q=u,d,\ldots$ in a bound proton originated by the energy-momentum distribution  and the off-shell effect.
Even using an isoscalar OS function $\delta f$, 
we observe a pronounced flavor dependence of the EMC effect at $x>0.5$ as a result of the 
convolution of PDFs with different $x$ dependence with the nucleon momentum distribution.
The nuclear dependence of $R_q$ is also noticeable in Fig.\ref{fig:RduA3} and is owed to the differences in the proton spectral functions of $\htri$ and $\hetri$. 
In order to further clarify the flavor dependence of nuclear effects, we show the asymmetry $\Delta_3=[R_q(\htri)-R_q(\hetri)]/[R_q(\htri)+R_q(\hetri)]$ for both $q=u$ and $q=d$ quarks in the right panel of Fig.~\ref{fig:RduA3}.
Our results contrast with those of Ref.~\cite{Cocuzza:2021rfn}, 
claiming a significant isovector nuclear EMC effect derived from a global QCD analysis including $A=2$ and $A=3$ DIS data (see Fig.~3 in \cite{Cocuzza:2021rfn}).
We comment in this context that Ref.~\cite{Cocuzza:2021rfn} uses the mHT model of HT terms in their analysis, which can introduce 
significant biases due to the interplay of the nucleon isospin dependence of the OS correction and the $d/u$ ratio and the HT terms. 
Furthermore, the analysis of Ref.~\cite{Cocuzza:2021rfn} 
introduces an explicit nuclear dependence in the 
OS functions for individual quark flavors, which 
may result in additional correlations among parameters potentially affecting the results.
We note that the HT terms cancel out in the ratio $F_i^n/F_i^p = F_i^{LT, n}/F_i^{LT, p}$ in the mHT model provided that $h_i^p=h_i^n$.
We therefore expect that analyses of the $^3$He/$^3$H SF ratio based on a naive LT approximation~\cite{Segarra:2021exb} 
could be affected by somewhat similar biases on the 
resulting isospin dependence of the OS function $\delta f$ as the ones found in the mHT model. 

\begin{figure}[htb]
\centering
\includegraphics[width=0.70\textwidth]{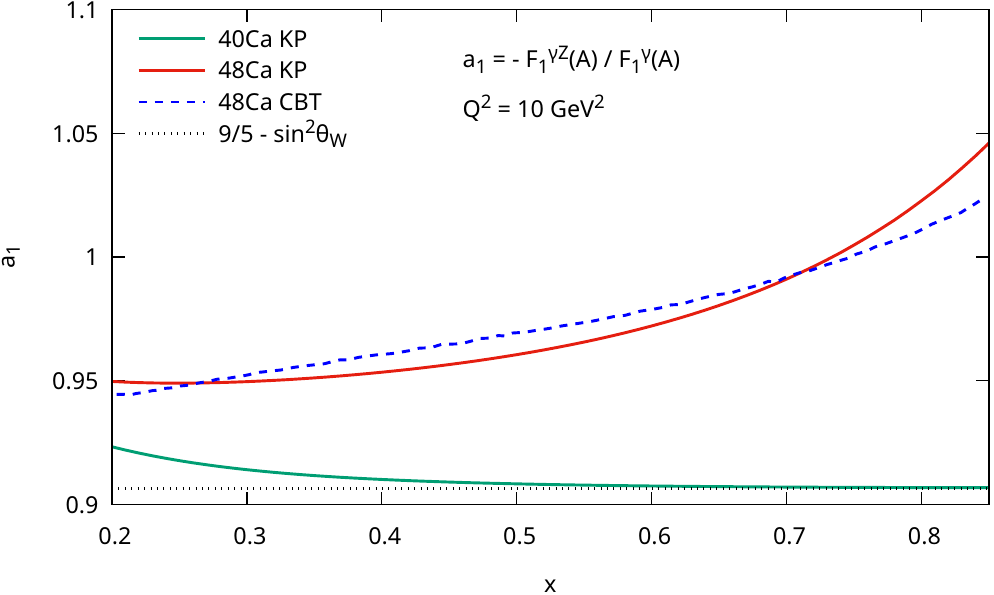}
\caption{%
Predictions of the KP model~\cite{Kulagin:2004ie} for the asymmetry $a_1=-F_1^{\gamma Z}/F_1^{\gamma}$ in parity-violating DIS off ${}^{48}$Ca (red solid) and ${}^{40}$Ca (green solid) targets at $Q^2=10\gevsq$. The CBT model~\cite{Cloet:2012td} result is taken from Refs.~\cite{Beminiwattha:2023est,Gaskell:PC}.
\label{fig:PVEMC}} 
\end{figure}
The isospin dependence in the nuclear modification of SFs can 
be probed in parity-violating DIS with the interference of 
the exchanged virtual photon and neutral $Z$ boson. 
In Fig.~\ref{fig:PVEMC} we show our predictions for the ratio 
$a_1=-F_1^{\gamma Z}/F_1^{\gamma}$ obtained with an isospin-symmetric function $\delta f^p=\delta f^n$ and the kinematics expected in the SoLID experiment at JLab~\cite{Beminiwattha:2023est}.
The comparison between ${}^{48}$Ca and the isoscalar ${}^{40}$Ca is instructive. We observe a significant enhancement on the asymmetry coefficient $a_1(x)$ in ${}^{48}$Ca even with an isospin-symmetric OS function $\delta f$, consistently with the $R_{u,d}$ nuclear 
modifications described above. 
It is of note that such an enhancement is comparable 
with the one expected in Ref.~\cite{Cloet:2012td} based on
a model with a strong isovector field. 
In this context, we therefore emphasize the importance 
of properly taking into account the flavor dependence arising 
from conventional nuclear effects. 
Precise measurements of the isospin dependence of nuclear effects at the parton level can be obtained from future flavor sensitive data from DIS at the electron-ion collider~\cite{AbdulKhalek:2021gbh} and from both neutrino and antineutrino charged-current interactions off hydrogen and various isoscalar and non-isoscalar nuclear 
targets~\cite{Petti:2019asx,Petti:2022bzt,Petti:2023osk} at the long-baseline neutrino facility~\cite{DUNE:2020ypp}. 

We thank G.~G.~Petratos for clarifications about the \mara{} data 
and G.~Salm\`e for providing the $\htri$ and $\hetri$ spectral functions of Ref.~\cite{Pace:2001cm}.


\begin{thebibliography}{99}

\bibitem{Alekhin:2022uwc}
S.~I.~Alekhin, S.~A.~Kulagin and R.~Petti,
Phys. Rev. D \textbf{107}, no.5, L051506 (2023)
[arXiv:2211.09514 [hep-ph]].

\bibitem{Alekhin:2022tip}
S.~I.~Alekhin, S.~A.~Kulagin and R.~Petti,
Phys. Rev. D \textbf{105}, no.11, 114037 (2022)
[arXiv:2203.07333 [hep-ph]].

\bibitem{JeffersonLabHallATritium:2021usd}
D.~Abrams \textit{et al.}, 
Phys. Rev. Lett. \textbf{128}, no.13, 132003 (2022)
[arXiv:2104.05850 [hep-ex]].

\bibitem{Alekhin:2017fpf}
S.~I.~Alekhin, S.~A.~Kulagin and R.~Petti,
Phys. Rev. D \textbf{96}, no.5, 054005 (2017)
[arXiv:1704.00204 [nucl-th]].

\bibitem{Kulagin:2004ie}
S.~A.~Kulagin and R.~Petti,
Nucl. Phys. A \textbf{765}, 126-187 (2006)
[arXiv:hep-ph/0412425].

\bibitem{Akulinichev:1985ij}
S.~V.~Akulinichev, S.~A.~Kulagin and G.~M.~Vagradov,
Phys. Lett. B \textbf{158}, 485 (1985).

\bibitem{Kulagin:1989mu}
S.~A.~Kulagin,
Nucl. Phys. A \textbf{500}, 653 (1989).

\bibitem{Kulagin:1994fz}
S.~A.~Kulagin, G.~Piller and W.~Weise,
Phys. Rev. C \textbf{50}, 1154 (1994)
[arXiv:nucl-th/9402015].

\bibitem{Georgi:1976ve}
H.~Georgi and H.~D.~Politzer,
Phys. Rev. D \textbf{14}, 1829 (1976).

\bibitem{Accardi:2016ndt}
A.~Accardi, \textit{et al.}, 
Eur. Phys. J. C \textbf{76}, no.8, 471 (2016)
[arXiv:1603.08906 [hep-ph]].

\bibitem{Virchaux:1991jc}
M.~Virchaux and A.~Milsztajn,
Phys. Lett. B \textbf{274}, 221 (1992).

\bibitem{Kulagin:2010gd}
S.~A.~Kulagin and R.~Petti,
Phys. Rev. C \textbf{82}, 054614 (2010)
[arXiv:1004.3062 [hep-ph]].

\bibitem{Kulagin:2014vsa}
S.~A.~Kulagin and R.~Petti,
Phys. Rev. C \textbf{90}, no.4, 045204 (2014)
[arXiv:1405.2529 [hep-ph]].

\bibitem{Kulagin:2007ju}
S.~A.~Kulagin and R.~Petti,
Phys. Rev. D \textbf{76}, 094023 (2007)
[arXiv:hep-ph/0703033].

\bibitem{Ru:2016wfx}
P.~Ru, S.~A.~Kulagin, R.~Petti and B.~W.~Zhang,
Phys. Rev. D \textbf{94}, no.11, 113013 (2016)
[arXiv:1608.06835 [nucl-th]].

\bibitem{Wiringa:1994wb}
R.~B.~Wiringa, V.~G.~J.~Stoks and R.~Schiavilla,
Phys. Rev. C \textbf{51}, 38 (1995)
[arXiv:nucl-th/9408016].

\bibitem{Veerasamy:2011ak}
S.~Veerasamy and W.~N.~Polyzou,
Phys. Rev. C \textbf{84}, 034003 (2011)
[arXiv:1106.1934 [nucl-th]].

\bibitem{Pace:2001cm}
E.~Pace, G.~Salme, S.~Scopetta and A.~Kievsky,
Phys. Rev. C \textbf{64}, 055203 (2001)
[arXiv:nucl-th/0109005].

\bibitem{Abrams:2024wgt}
D.~Abrams \textit{et al.}, 
arXiv:2410.12099 [nucl-ex].

\bibitem{Seely:2009gt}
J.~Seely \textit{et al.}, 
Phys. Rev. Lett. \textbf{103}, 202301 (2009)
[arXiv:0904.4448 [nucl-ex]].

\bibitem{HERMES:1999bwb}
K.~Ackerstaff \textit{et al.}, 
Phys. Lett. B \textbf{475}, 386-394 (2000)
[erratum: Phys. Lett. B \textbf{567}, 339-346 (2003)]
[arXiv:hep-ex/9910071 [hep-ex]].

\bibitem{Cocuzza:2021rfn}
C.~Cocuzza \textit{et al.}, 
Phys. Rev. Lett. \textbf{127}, no.24, 242001 (2021)
[arXiv:2104.06946 [hep-ph]].

\bibitem{Beminiwattha:2023est}
R.~Beminiwattha, J.~Arrington and D.~J.~Gaskell,
Eur. Phys. J. A \textbf{59}, no.8, 194 (2023)
[arXiv:2304.04622 [nucl-ex]].

\bibitem{Gaskell:PC}
D.~Gaskell, private communications (2021).

\bibitem{Segarra:2021exb}
E.~P.~Segarra, \textit{et al.}, 
arXiv:2104.07130 [hep-ph].

\bibitem{Cloet:2012td}
I.~C.~Cloet, W.~Bentz and A.~W.~Thomas,
Phys. Rev. Lett. \textbf{109}, 182301 (2012)
[arXiv:1202.6401 [nucl-th]].

\bibitem{AbdulKhalek:2021gbh}
R.~Abdul Khalek, \textit{et al.}, 
Nucl. Phys. A \textbf{1026}, 122447 (2022)
[arXiv:2103.05419 [physics.ins-det]].

\bibitem{Petti:2019asx}
R.~Petti,
arXiv:1910.05995 [hep-ex].

\bibitem{Petti:2022bzt}
R.~Petti,
Phys. Lett. B \textbf{834}, 137469 (2022)
[arXiv:2205.10396 [hep-ph]].

\bibitem{Petti:2023osk}
R.~Petti,
arXiv:2301.04744 [hep-ex].

\bibitem{DUNE:2020ypp}
B.~Abi \textit{et al.}, 
arXiv:2002.03005 [hep-ex].


\end{thebibliography}
\end{document}